\documentclass[onecolumn,amsmath,showpacs,nofootinbib,11pt]{revtex4-2}
\usepackage{graphicx}
\usepackage{dcolumn}
\usepackage{bm}

\begin{document}

\title{Brane-vector dark matter and its connection to inflation and primordial gravitational waves}


\author{Cao H. Nam}
\email{nam.caohoang@phenikaa-uni.edu.vn (corresponding author)} 
\affiliation{Phenikaa Institute for Advanced Study and Faculty of Fundamental Sciences, Phenikaa University, Yen Nghia, Ha Dong, Hanoi 12116, Vietnam}
\author{Tran N. Hung}
\email{hung.tranngoc@phenikaa-uni.edu.vn}  
\affiliation{Phenikaa Institute for Advanced Study and Faculty of Fundamental Sciences, Phenikaa University, Yen Nghia, Ha Dong, Hanoi 12116, Vietnam}
\date{\today}

\begin{abstract}%
The scalar mode describing the fluctuation of the 3-brane (the observable universe) in a five-dimensional bulk spacetime compactified on a circle is absorbed by the Kaluza-Klein U(1) gauge field, leading to a massive brane-vector living on the 3-brane. The brane-vector can be responsible for dark matter because it is odd under a $\mathrm{Z}_2$ symmetry, neutral under the Standard Model (SM) symmetries, and couples extremely weak to the SM particles due to its gravitational origin. Interestingly, the brane-vector dark matter could leave particular imprints on the cosmic microwave background (CMB) and the primordial gravitational waves. Hence, the precise measurements of the CMB and the observations of the primordial gravitational waves generated during the inflation can provide a potential way to probe the extra-dimensions and branes which are the main ingredients of string/M theory.

\end{abstract}

\maketitle
\section{Introduction}

Extra-dimensions and branes (hypersurfaces embedded in a higher-dimensional spacetime or the bulk) play a fundamental role in string/M theory which has been considered as a promising candidate for quantum gravity. String/M theory predicts that the spacetime has spatial extra-dimensions wrapped upon a compact internal space and open strings describing the non-gravitational sector with their endpoints are attached to branes. Whereas, closed strings describing the gravitational sector can propagate freely in the bulk. The low energy regime of string/M theory hence can imply a scenario that our observable universe is actually a 3-brane on which the Standard Model (SM) fields are confined, the directions transverse to the 3-brane are compact, and only the gravitational degrees of freedom and other exotic ones can move freely in the bulk \cite{Horava1996,Lukas1999}. In such a scenario, the constraints on the size of the extra-dimensions could be obtained from deviations from the usual Newton's law \cite{Adelberger2003,Adelberger2009,Kapner2007,Murata2015} and the experiments of gravitational waves (GW) \cite{Trashorras2016,Visinelli2018,SenGupta2018}. Therefore, the experimental bounds of the size of the extra-dimensions can be relaxed, where the size of the extra-dimensions is as large as the micron scale. Interestingly, the compactified extra-dimensions can provide a possible solution to the hierarchy problem where the fundamental Planck scale can be much lower than the observed Planck scale \cite{Dvali1998,Nam2021}.

The brane tension is an important parameter that characterizes the typical size of the brane fluctuations along the extra-dimensions. If the brane tension is large enough, corresponding to the branes being too heavy, the branes can be considered as rigid objects. On the contrary, when the brane tension is much lower than the fundamental scale, the branes can fluctuate along the extra-dimensions. The fluctuations corresponding to the oscillating brane are realized as the Nambu-Goldstone bosons, also well-known as the branons, which arise as a result of the spontaneously broken isometry of the internal space due to the presence of the brane \cite{Sundrum1999}. The branons are the physical scalar excitations in the low energy spectrum of the effective field theory generically only in the global case where the physical degrees of freedom describing the higher-dimensional gravitation are negligible. 

From the Kaluza-Klein (KK) theory, we know that the isometry of the internal space leads to the gauge transformations where the gauge fields are the off-diagonal components of the bulk metric \cite{Bailin1987,Overduin1997} or the connection on the spacetime fiber bundle with the typical fiber being the internal manifold \cite{Coquereaux1998,Nam2019}. This implies when the isometry of the internal space is broken spontaneously by the presence of the brane, the KK gauge fields would absorb the branons as their longitudinal modes, corresponding to a gravitational Brout-Englert-Higgs mechanism. This leads to massive vector fields appearing on the brane, as well-known as brane-vector \cite{Dobado2001,Kang2001,Clark2007,Clark2008,Clark2009}. The mass of the brane-vector is determined by the brane tension, and hence it is a free parameter. Because the brane tension is much lower than the fundamental Planck scale for the case of the flexible branes, we expect that the mass of the brane-vector can range from the eV regime to the TeV regime. The brane-vector interacts with the SM fields via their energy-momentum tensor with the coupling determined by the observed Planck scale as a result of the fact that the brane-vector has a gravitational origin as new dynamic degrees of freedom describing the behavior of the spacetime at the short-distances \cite{Nam2023}.

In this paper, we explore the brane-vector as a candidate for dark matter (DM) whose stability is guaranteed by
a $\mathrm{Z}_2$ parity as a result of the 4D general coordinate invariance on the brane. The coupling of the brane-vector to the SM particles is strongly suppressed by the observed Planck scale, hence the relic abundance of the brane-vector DM could not generated by the thermal processes that appeared after the inflation. On the other hand, the brane-vector DM which couples extremely weak to the SM particles could be produced by the gravitational mechanism due to the quantum fluctuations in the inflationary stage \cite{Ford1987,Lyth1998,Dimopoulos2006,Arias2012,Graham2016}. The transverse polarizations of a massive vector field that couples minimally to 4D gravity are conformally invariant in analogy to the massless vector field, since its production is suppressed and could be negligible during the inflation \cite{Arias2012,Graham2016}. In this sense, the observed relic abundance of the brane-vector DM should result from the production of its longitudinal polarization.

The Friedmann equation describing the evolution of the brane universe consists of linear and quadratic terms in terms of the energy density. The quadratic term is inversely proportional to the mass of the brane-vector or the brane tension since it plays an important role in the early universe. (Whereas, the linear term is dominant when the energy density is much smaller than the brane tension corresponding to the late time of the universe evolution.) This novel feature of the modified Friedmann equation combined with the production of the brane-vector DM during the inflation leads to a prediction of the tensor-to-scalar ratio and the reheating temperature in terms of the brane-vector mass. Note that, the reheating temperature can be estimated by the primordial GW spectrum \cite{Nakayama2008}. In this way, the brane-vector DM could leave specific imprints on the cosmic microwave background (CMB) and the primordial GWs. Therefore, the precise measurements of the CMB as well as the future detections of the primordial GWs and the B-mode polarization in the CMB can allow us to probe the extra-dimensions and branes.

This paper is organized as follows. In Sec. \ref{BVDM}, we study the four-dimensional effective field theory derived from a flexible $3$-brane which propagates in a bulk spacetime compactified on a circle $S^1$ and on which the SM and inflaton fields are confined. We show that the low-energy spectrum of the effective theory includes a brane-vector that possesses a discrete $\mathrm{Z}_2$ parity and hence could be responsible for the DM. Then, we compute the relic abundance of the brane-vector DM based on the production of its longitudinal mode during the inflation. In Sec. \ref{Connect}, we study the imprints of the brane-vector on the spectrum of the CMB and the primordial GWs carrying the information of the inflation era for two typical potentials of the inflation: i) $\alpha$-attractor potential and ii) natural potential. Finally, we present a brief conclusion in Sec. \ref{conclu}.

\section{\label{BVDM} Brane-vector DM}

For simplicity without loss of generality, we consider the dynamics of a $3$-brane propagating in a five-dimensional bulk spacetime compactified on a circle $S^1$. The action of the system is given by a sum of the bulk action and the brane action as follows
\begin{eqnarray}
S&=&S_{\text{bulk}}+S_{\text{brane}},\label{totact}\\
S_{\text{bulk}}&=&\frac{M^3_5}{2}\int d^5X\sqrt{-G}\left(\mathcal{R}^{(5)}-2\Lambda_5\right),\label{bulkact}\\
S_{\text{brane}}&=&\int d^4x\sqrt{-\widetilde{g}}\left[-f^4+\mathcal{L}_{\text{SM}}+\frac{1}{2}\widetilde{g}^{\mu\nu}\partial_\mu\phi\partial_\nu\phi-V(\phi)\right].\label{braneact}
\end{eqnarray}
In the bulk action (\ref{bulkact}), $M_5$ is the five-dimensional Planck scale, $\mathcal{R}^{(5)}$ is the Ricci scalar of the bulk spacetime, $G$ is the determinant of the bulk metric, $X^M$ with $M=0,1,2,3,4$ are the coordinates of the points on the bulk spacetime, and $\Lambda_5$ refers to the bulk cosmological constant which is necessarily negative to restore the conventionally Hubble expansion law in the low-energy regime \cite{Clifton2012}. In the brane action (\ref{braneact}), $f$ is the tension of the $3$-brane, the coordinates of the points on the $3$-brane are parameterized by $x^\mu$ with $\mu=0,1,2,3$, and $\widetilde{g}$ is the determinant of the induced metric $\widetilde{g}_{\mu\nu}$ on the $3$-brane. The SM fields that are described by the Lagrangian $\mathcal{L}_{\text{SM}}$ and the inflaton field $\phi$ are confined to the $3$-brane. 

The presence of the $3$-brane could be created spontaneously by a proper physical process, such as the tachyon condensation when turning on the magnetic fluxes \cite{Majumdar2002}. This configuration should be realized as the ground state. As a result, the formation of the $3$-brane would break spontaneously the local isometry of the bulk spacetime in the compactified extra dimension. As we see below, the Nambu-Goldstone (NG) mode describing the excitation of the $3$-brane would be eaten by the KK gauge field, corresponding to the local $U(1)$ isometry of the circle $S^1$, hence it becomes massive. 

\subsection{Dimensional reduction of the bulk action}

Let us study the compactification of the five-dimensional bulk spacetime on the circle $S^1$ that is parametrized by an angle $\theta$. The bulk metric can be decomposed into the four-dimensional fields as follows
\begin{eqnarray}
G_{MN}=\left(%
\begin{array}{cc}
  g_{\mu\nu}+\rho^2 A_\mu A_\nu & \rho^2 A_\mu \\
  \rho^2 A_\nu & \rho^2 \\
\end{array}%
\right),\label{bulkmetr}
\end{eqnarray}
where $g_{\mu\nu}$ is the four-dimensional metric tensor, $A_\mu$ is the KK gauge field corresponding to the local isometry in the fifth compactified dimension, and $\rho$ is a scalar field whose vacuum expectation value (vev) determines physically the radius of the circle $S^1$. With this decomposition, we can write the bulk action (\ref{bulkact}) in terms of the four-dimensional fields as follows.
\begin{eqnarray}
S_{\text{bulk}}&=&\frac{M^2_{\text{P}}}{2\langle\rho\rangle}\int d^4x\sqrt{-g}\rho\left(\mathcal{R}^{(4)}-2\Lambda_5-\frac{\rho^2}{4}F_{\mu\nu}F^{\mu\nu}+\frac{2}{3}\frac{\partial^\mu\rho\partial_\mu\rho}{\rho^2}\right),\label{KKact}
\end{eqnarray}
where $g$ is the determinant of the four-dimensional metric $g_{\mu\nu}$, $\mathcal{R}^{(4)}$ is the four-dimensional Ricci scalar, $F_{\mu\nu}\equiv\partial_\mu A_\nu-\partial_\nu A_\mu$ is the field strength tensor of the KK gauge field, and $M_{\text{P}}$ is the observed Planck scale determined as $M^2_{\text{P}}=2\pi\langle\rho\rangle M^3_5$ with $\langle\rho\rangle$ denoting the vev of the scalar field $\rho$ or the radius of the circle $S^1$. We can check that the gravitational action (\ref{KKact}) is invariant under the following transformation 
\begin{eqnarray}
\theta&\longrightarrow&\theta+\alpha(x),\\
A_\mu&\longrightarrow&A_\mu-\partial_\mu\alpha(x).
\end{eqnarray}
A set of such transformations forms the local isometry in the compactified dimension associated with a particular class of general coordinate transformations. In particular, the off-diagonal element of the bulk metric transforms as a gauge field of the U(1) local symmetry.

Reducing the bulk action (\ref{bulkact}) on the internal manifold $S^1$, we obtain an effective action that consists of the four-dimensional Einstein gravity plus a cosmological constant coupled to the massless U(1) gauge field and the scalar field as follows
\begin{eqnarray}
S_{4D}&\supset&\int d^4x\sqrt{g}\left[\frac{M^2_{\text{P}}}{2}\left(\mathcal{R}^{(4)}-2\Lambda_5\right)-\frac{1}{4\kappa^2}F_{\mu\nu}F^{\mu\nu}+\frac{1}{2}\partial^\mu r\partial_\mu r+\cdots\right],\label{4Dgravact}
\end{eqnarray}
where we have expanded the scalar field $\rho$ around its vev as $\rho=\langle\rho\rangle(1+\sqrt{3}r/2)$ with $r$ to be the radion field, $\kappa$ is the KK gauge coupling given as 
\begin{eqnarray}
\kappa\equiv\sqrt{2}\frac{\langle\rho\rangle^{-1}}{M_{\text{P}}},\label{KKcoup}
\end{eqnarray}
and the ellipse refers to the terms related to the self-couplings of the radion field and its couplings to the four-dimensional metric and the KK gauge field.

\subsection{Effective action on the $3$-brane}

The induced metric on the $3$-brane can be determined by the embedding geometry as $ds^2_{\text{brane}}=\widetilde{g}_{\mu\nu}dx^\mu dx^\nu$ where $\widetilde{g}_{\mu\nu}=G_{MN}\partial_\mu Y^M\partial_\nu Y^N$ with $Y^M(x)$ being the embedding function which describes the position of the $3$-brane in the bulk spacetime. In the static gauge, the embedding function is given by $Y^M(x)=(x^\mu,Y^\theta(x))$. With the ansatz of the bulk metric as given by Eq. (\ref{bulkmetr}), the induced metric $\widetilde{g}_{\mu\nu}$ is found as follows
\begin{eqnarray}
\widetilde{g}_{\mu\nu}&=&g_{\mu\nu}-\rho^2\left(A_\mu+\partial_\mu Y^\theta\right)\left(A_\nu+\partial_\nu Y^\theta\right),\nonumber\\
&=&g_{\mu\nu}-\langle\rho\rangle^2\left(A_\mu+\partial_\mu Y^\theta\right)\left(A_\nu+\partial_\nu Y^\theta\right)+\mathcal{O}(r).\label{indmetr}
\end{eqnarray}

When the $3$-brane is created at a certain point $Y^\theta_0$ in the fifth compactified dimension which minimizes the brane action (\ref{braneact}) and corresponds to the ground state. In this situation, the presence of the $3$-brane breaks spontaneously the U(1) local isometry in the fifth compactified dimension. The fluctuations of the $3$-brane around the ground state in the fifth compactified dimension can be parametrized as $Y^\theta(x)=Y^\theta_0+\pi(x)$ where $\pi(x)$ is the NG mode associated with the excitation of the $3$-brane along the direction of the U(1) local isometry breaking. As a result, the KK gauge field would absorb this NG mode as its longitudinal mode, leading to the brane-vector. This brane-vector is expressed by the definition $X_\mu\equiv A_\mu+\partial_\mu\pi$. The field strength tensor of the brane-vector is given by 
\begin{eqnarray}
X_{\mu\nu}\equiv\partial_\mu X_\nu-\partial_\nu X_\mu=F_{\mu\nu}.    
\end{eqnarray}

It should be noted that the NG mode describing the fluctuation of the $3$-brane in the fifth compactified dimension would be the physical excitation in the decoupling limit of the KK gauge field corresponding to take the limit $\kappa\rightarrow0$. This physical excitation is well known as the branon which is massive in the situation that the U(1) global isometry in the fifth compactified dimension is realized as an approximate symmetry. Interestingly, it was indicated that the branon can be responsible for the DM candidate \cite{Cembranos2003}. However, as indicated in Ref. \cite{Nam2023}, this limit leads to the violation of the Sublattice Weak Gravity Conjecture \cite{Reece2017,Reece2018}, the Swampland Distance Conjecture \cite{Ooguri2007}, and the Festina Lente bound \cite{Montero2020,Vafa2021} which are recognized as the constraints of quantum gravity on the low-energy effective theories. On the other hand, this limit would belong to the Swampland which consists of the effective field theories that cannot be ultraviolet completed in quantum gravity. 

In this way, we can rewrite the induced metric on the $3$-brane in terms of the four-dimensional metric and the brane-vector as follows
\begin{eqnarray}
\widetilde{g}_{\mu\nu}=g_{\mu\nu}-\langle\rho\rangle^2X_\mu X_\nu+\mathcal{O}(r).\label{indmetr2}
\end{eqnarray}
Using this result, we can expand the square root of the determinant of the induced metric on the $3$-brane as follows
\begin{eqnarray}
\sqrt{-\widetilde{g}}=\sqrt{-g}\left(1-\frac{\langle\rho\rangle^2}{2}X^\mu X_\mu+\cdots\right),\label{indmetr-det}
\end{eqnarray}
where $g$ refers to the determinant of the four-dimensional metric $g_{\mu\nu}$. With Eqs. (\ref{indmetr2}) and (\ref{indmetr-det}), we can find an effective action on the $3$-brane as follows
\begin{eqnarray}
S_{4D}&\supset&\int d^4x\sqrt{-g}\left(-f^4+\frac{m^2_X}{2\kappa^2}X_\mu X^\mu+\frac{\lambda}{\kappa^2}X^\mu X^\nu T^{\text{mt}}_{\mu\nu}+\mathcal{L}_{\text{SM}}+\frac{1}{2}g^{\mu\nu}\partial_\mu\phi\partial_\nu\phi-V(\phi)+\cdots\right),\label{effbract}
\end{eqnarray}
where $m_X=\sqrt{2}f^2/M_{\text{P}}$ is the mass of the brane-vector which is acquired due to the spontaneous breaking of the U(1) local isometry in the fifth compactified dimension, $T^{\text{mt}}_{\mu\nu}$ is the energy-momentum tensor of the matte fields (the SM and inflation fields) which is given by
\begin{eqnarray}
T^{\text{mt}}_{\mu\nu}=2\frac{\delta\mathcal{L}_{\text{mt}}}{\delta g^{\mu\nu}}-g_{\mu\nu}\mathcal{L}_{\text{mt}},
\end{eqnarray}
with $\mathcal{L}_{\text{mt}}$ denote the Lagrangian of the matter fields living on the $3$-brane, and $\lambda=1/M^2_{\text{P}}$ is the coupling constant of the brane-vector to the matter fields.

From the actions (\ref{4Dgravact}) and (\ref{effbract}), we observe that first, the brane-vector is odd under a discrete symmetry $\mathrm{Z}_2$. This means that the effective action (\ref{effbract}) is invariant under the transformation $X_\mu\rightarrow-X_\mu$, which means that the brane-vector always appears in the pairs. Hence, this discrete symmetry would guarantee the stability of the brane vector. Second, in analogy to the graviton, the brane-vector couples to the energy-momentum tensor of the matter fields with the coupling strength suppressed strongly by the observed Planck scale. This is due to the fact that the brane vector has a geometric or gravitational origin associated with the off-diagonal component of the bulk metric. Because the brane-vector is stable, neutral, and couples very weakly to the matter fields, it behaves as the DM. 

\subsection{Production of brane-vector particles during the inflation}

Because the brane-vector is a massive vector field, it has three physical polarizations corresponding to two transverse components and one longitudinal component which is the NB mode describing the fluctuation of the $3$-brane in the fifth compactified dimension. However, the production of the transverse polarizations of the brane-vector by the quantum fluctuations is suppressed during the inflation because the transverse polarizations are conformally invariant. This implies that the contribution of the transverse polarizations to the production of the energy density of the brane-vector can be negligible. Whereas, unlike transverse polarizations, the production of the longitudinal polarization is unsuppressed because it is not conformally invariant. Therefore, the relic abundance of the brane-vector DM which is observed today is essentially due to the production of its longitudinal polarization during the inflation.

The relic abundance of the brane-vector observed today is
\begin{eqnarray}
\Omega_X=\frac{\rho_X}{\rho_c},
\end{eqnarray}
where $\rho_c$ is the critical density and $\rho_X$ is the present energy density of the brane-vector. As discussed above, the contribution to the present energy density of the brane-vector is approximately that of its longitudinal mode and it is calculated as follows \cite{Graham2016}
\begin{eqnarray}
\rho_X\simeq\rho_{X,L}=\frac{H^2_i}{8\pi^2}\frac{a_*k^2_*}{a^3_0}\int d\ln k\frac{k^2}{k^2_*}\left[\frac{a_0}{a_*}\left|\frac{X_L(k,t_0)}{X_0(k)}\right|^2+\frac{a^3_0}{a_*\left(k^2+a^2_0m^2_X\right)}\left|\frac{\partial_tX_L(k,t_0)}{X_0(k)}\right|^2\right],\label{Bra-rel-abun}
\end{eqnarray}
where $H_i$ refers to the Hubble parameter during the inflation, $a_0$ and $a_*$ are the values of the scale factor at the present time and the moment of $H=m_X$, respectively, the wavenumber $k_*$ is defined as $k_*=a_*m_X$. $X_L(k,t_0)$ is the solution of the equation of motion for the longitudinal polarization of the brane-vector at the present time, given by
\begin{eqnarray}
\left(\partial^2_t+\frac{3k^2+a^2m^2_X}{k^2+a^2m^2_X}H\partial_t+\frac{k^2}{a^2}+m^2_X\right)X_L(k,t)=0,\label{XLequa}   
\end{eqnarray}
with an initial condition at the horizon exit given as
\begin{eqnarray}
X_L(k,t)\Big|_{\text{exit}}=X_0(k)\left(1-\frac{ik}{aH_i}\right)\exp\left\{\frac{ik}{aH_i}\right\}.
\end{eqnarray}
By solving numerically Eq. (\ref{XLequa}), we can find $X_L(k,t)$ and then substitute it into the integral given in Eq. (\ref{Bra-rel-abun}) to obtain an approximate result which is equal to $1.2$ \cite{Graham2016}. As a result, we derive the energy density of the brane-vector observed at the present time as follows
\begin{eqnarray}
\rho_X\simeq\frac{3m^2_XH^2_i}{20\pi^2}\left(\frac{a_*}{a_0}\right)^3.\label{rhoBV}
\end{eqnarray}
From the expansion law along the $3$-brane as \cite{Binetruy2000,Deffayet2000}
\begin{eqnarray}
H^2+\frac{k^2}{a^2}=\left(\frac{\rho_{\text{m}}+M^2_{\text{P}}m^2_X/2}{6M^3_5}\right)^2+\frac{\Lambda_5}{6},\label{NSHubb}
\end{eqnarray}
and the relation $a_*\simeq a_{\text{mre}}\sqrt{H_{\text{mre}}/m_X}$ with $a_{\text{mre}}$ and $H_{\text{mre}}$ denoting the scale factor and the Hubble parameter at the time of the matter-radiation equality, we can determine the ratio $a_*/a_0$ in Eq. (\ref{rhoBV}) as
\begin{eqnarray}
\frac{a_*}{a_0}=\left(\frac{\rho_{\text{r},0}}{(\sqrt{7}-1)M^2_{\text{P}}}\right)^{1/4}\frac{1}{\sqrt{m_X}},
\end{eqnarray}
where $\rho_{\text{r},0}$ is the energy density of the radiation at the present time. In addition, using the tensor power spectrum amplitude in the brane-world scenario as $A_t\simeq12x_0\left[H_i/(2\pi M_{\text{P}})\right]^2$ with $x_0\equiv2\sqrt{3H^2_i M^2_{\text{P}}/(2f^4)}$ \cite{Langlois2000} and the relation between the tensor power spectrum amplitude $A_t$ and the scalar power spectrum amplitude $A_s$ given by the tensor-to-scalar ratio $r$ as $r=A_t/A_s$, we can determine the Hubble parameter $H_i$ in terms of the inflationary quantities that can be observed in the CMB spectrum as
\begin{eqnarray}
 H^3_i=\frac{\pi^2A_sr}{6\sqrt{3}}m_XM^2_{\text{P}}.   
\end{eqnarray}
Finally, we find an expression for the relic abundance of the brane-vector DM in terms of its mass, the tensor-to-scalar ratio and the scalar power spectrum amplitude as
\begin{eqnarray}
\Omega_Xh^2&\simeq&0.16\times\left(\frac{rA_sM^2_{\text{P}}}{10^{26}\text{GeV}^2}\right)^{2/3}\left(\frac{m_X}{10^3 \text{GeV}}\right)^{7/6}.\label{KKGBDM-rab}
\end{eqnarray}

\begin{figure}[t]
\centering
\begin{tabular}{cc}
\includegraphics[width=0.6 \textwidth]{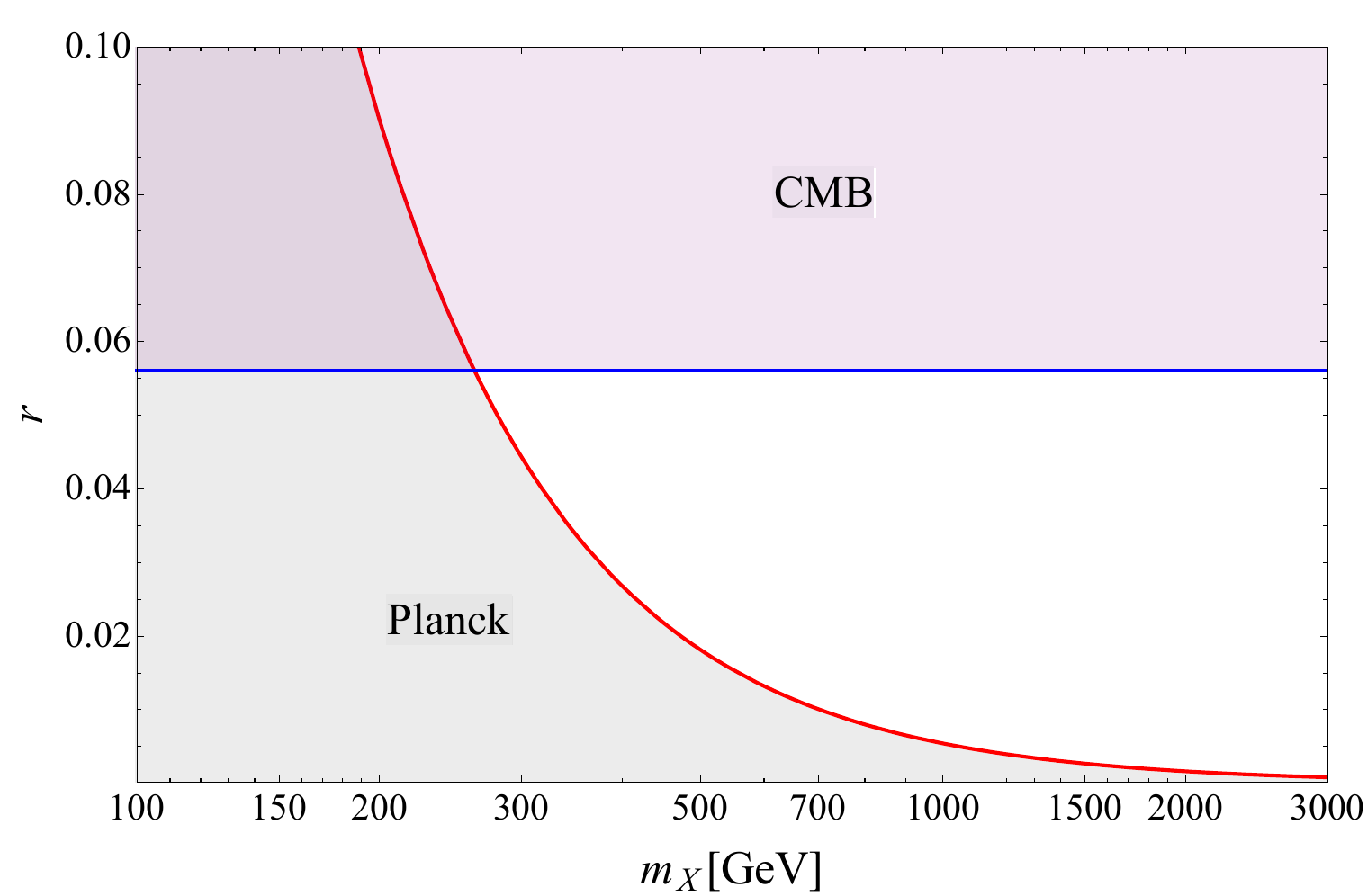}
\end{tabular}
 \caption{The allowed parameter region (the white region) is given in the $r-m_X$ plane where the light gray and light purple regions are excluded by the measurement of the DM relic density by the Planck collaboration \cite{Planck2020} and no detection of the CMB B-mode polarization leading to the $95\%$ CL upper limit on $r$ \cite{Planck2020b}, respectively.}\label{r-mX-rel}
\end{figure}

Using the constraints from both the Planck data on the observed DM relic abundance $\Omega_{\text{CDM}}h^2\simeq0.12$ \cite{Planck2020} and the $95\%$ CL upper limit on the tensor-to-scalar ratio as $r<0.056$ \cite{Planck2020b}, we depict the allowed parameter space for the tensor-to-scalar ratio $r$ and the brane-vector mass $m_X$ in Fig. \ref{r-mX-rel}.
We observe that these constraints impose a lower bound on the brane-vector mass as $m_X\gtrsim262.45$ GeV. This bound can be transferred to an upper bound on the radius of the compactified extra-dimension. First, we need to determine the brane-vector mass in terms of the radius of the compactified extra-dimension. We observe that the expansion law on the $3$-brane as given by Eq. (\ref{NSHubb}) would lead to the standard expansion law in the low energy limit of $\rho_{\text{m}}\ll M^2_{\text{P}}m^2_X/2$ and the bulk cosmological constant satisfying 
\begin{eqnarray}
\Lambda_5=-\frac{6M^6_5}{M^4_{\text{P}}}.    
\end{eqnarray}
Using this relation, we find an expression for the brane-vector mass in terms of $\langle\rho\rangle$ as follows
\begin{eqnarray}
m_X=\frac{\sqrt{3}}{\pi\langle\rho\rangle}.  
\end{eqnarray}
Finally, we determine the constraint on the radius of the compactified extra-dimension as $\langle\rho\rangle\lesssim4.14\times10^{-19}$ m in order for the brane-vector responsible for the DM candidate. Obviously, this upper bound satisfies the constraint coming from the fifth-force experiments \cite{Kapner2007,Murata2015} which set an upper bound $\langle\rho\rangle\lesssim \mathcal{O}(40)$ $\mu$m. In addition, we find that the brane-vector mass is about a few hundred GeV for the region $r\in[\mathcal{O}(10^{-3}),\mathcal{O}(10^{-2})]$ which is predicted in some potential inflation models and is experimentally accessible in the near future/present experiments.

\section{\label{Connect} Connection of brane-vector DM to inflation and primordial gravitational wave}
As discussed above, the brane-vector DM has the geometric origin by which its coupling to the SM particles would be strongly suppressed by the observed Planck scale. Hence, it is extremely hard to detect the signatures of the brane-vector DM at the colliders. However, the inflation and the primordial GWs generated during the inflation are very sensitive to the presence of the brane-vector DM. This can be seen from the non-standard expansion law given by Eq. (\ref{NSHubb}) with the allowed mass region of $m_X\gtrsim262.45$ GeV. The very different behavior of the expansion rate compared to the usual one due to the presence of the brane-vector DM becomes significant in the energy density region around/above $5.32\times10^{39}$ GeV$^4$. Therefore, the production of the brane-vector DM due to the quantum fluctuations during the inflation would leave imprints in the spectrum of the CMB and the primordial GWs whose accurate measurements allow us to probe the brane-vector DM and hence the compactified extra dimension and the brane universe. 

In the following, we assume that the brane-vector particles constitute almost the observed relic abundance by which the tensor-to-scalar ratio can be expressed in terms of the brane-vector mass as follows
\begin{eqnarray}
r\simeq\left(\frac{50.55\ \text{GeV}}{m_X}\right)^{7/4}.\label{rmX-rel}    
\end{eqnarray}
From this relation, we observe that if the tensor-to-scalar ratio is not too small, the energy density during the inflation is much larger than $M^2_{\text{P}}m^2_X/2$. As a result, the expansion law during the inflation is given by
\begin{eqnarray}
H_i\simeq\frac{V(\phi)}{\sqrt{3}M^2_{\text{P}}m_X}.
\end{eqnarray}
The equation of motion which plays the behavior of the inflaton during the inflation in the slow-roll approximation reads
\begin{eqnarray}
3H_i\dot{\phi}\simeq-V'(\phi).
\end{eqnarray}
Also, we can determine the slow-roll parameters in terms of the inflation potential, its derivatives, and the brane-vector mass as follows
\begin{eqnarray}
\epsilon\simeq\frac{M^4_{\text{P}}m^2_X}{V(\phi)}\left( \frac{V'(\phi)}{V(\phi)}\right)^2, \ \ \eta\simeq\frac{M^4_{\text{P}}m_X^2}{V(\phi)}\frac{V''(\phi)}{V(\phi)}.\label{sl-rol-pars}
\end{eqnarray}
These slow-roll parameters are related to the scalar spectral index $n_s$ and the tensor-to-scalar ratio $r$ which can be measured in the CMB spectrum as
\begin{eqnarray}
n_s-1=-6\epsilon+2\eta, \ \ r=16\epsilon.    
\end{eqnarray}
The number of e-folds between the moment that the comoving mode with the wave number $k$ exits the horizon and the end of the inflation is computed as 
\begin{eqnarray}
N_k&=&\int^{\phi_{\text{e}}}_{\phi_k}\frac{H_i}{\dot{\phi}}d\phi\nonumber\\
&=&-\frac{1}{M^4_{\text{P}}m^2_X}\int^{\phi_{\text{e}}}_{\phi_k}\frac{V^2(\phi)}{V'(\phi)}d\phi,\label{efoldNum}    
\end{eqnarray}
where $\phi_k$ and $\phi_{\text{e}}$ refer to the values of the inflaton at the time of the mode $k$ exiting the horizon and the end of the inflation, respectively. We can determine $\phi_{\text{e}}$ from $\epsilon\approx1$ which occurs at the end of the inflation. 

With the equations related to the inflation and the brane-vector DM production during the inflation, we can predict the scalar spectral index $n_s$ and the tensor-to-scalar ratio $r$ in terms of the number of e-folds $N_k$. The relation between $r$ and $N_k$ is given by
\begin{equation}
\frac{r^{15/7}}{1078.47}=\frac{M^4_{\text{P}}\text{GeV}^2}{V(\phi_k)}\left( \frac{V'(\phi_k)}{V(\phi_k)}\right)^2,
\end{equation}
where $\phi_k=\phi_k(N_k,r)$ which is a function of the number of e-folds and the tensor-to-scalar ratio and can be determined from Eqs. (\ref{rmX-rel}) and (\ref{efoldNum}).
Whereas, the relation between $n_s$ and $N_k$ is determined by 
\begin{eqnarray}
n_s=1-\frac{3}{8}r(N_k)+2\eta(N_k).
\end{eqnarray}

\subsection{$\alpha$-attractor potential}
The inflation with the $\alpha$-attractor potential is given by \cite{Carrasco2015a,Carrasco2015b}
\begin{eqnarray}
V(\phi)=\Lambda^{4}\left[ 1-\exp\left( -\frac{\phi}{\bar{M}}\right) \right] ^{2n}, \label{eqn:attpot}
\end{eqnarray}
where $\phi$ is the inflaton field coupled minimally to gravity on the 3-brane, $\Lambda$ is related to the inflationary energy scale, $\bar{M}\equiv\sqrt{3\alpha/2}M_{\text{P}}$ with $\alpha$ to be a parameter determining the inverse curvature of the K\"{a}hler manifold, and $n$ is a parameter. Note that, in the case of $\alpha=1$ and $n=1$, the $\alpha$-attractor potential leads to the Starobinsky model \cite{Starobinsky1980}.

By substituting the potential (\ref{eqn:attpot}) into Eq. (\ref{sl-rol-pars}), we obtain the slow-roll parameters as follows
\begin{eqnarray}
\epsilon&=&\left[\frac{2n\beta e^{-\phi/\bar{M}}}{\left(1-e^{-\phi/\bar{M}}\right)^{n+1}}\right]^{2}\label{eqn:eps},\\
\eta&=&\left(1-\frac{e^{\phi/\bar{M}}}{2n}\right)\epsilon,
\end{eqnarray}
where
\begin{eqnarray}
\beta\equiv\frac{M_{\text{P}}^{2}m_{X}}{\bar{M}\Lambda^{2}}.
\end{eqnarray}
At the moment that the mode $k$ crosses the comoving Hubble horizon corresponding to $X_{k}=\exp\left(-\phi_k/\bar{M}\right)\ll 1$, we obtain the following approximation
\begin{eqnarray}
X_k&\simeq&\frac{\sqrt{\epsilon}}{2n\beta},\\
\eta&\simeq&\epsilon-\beta\sqrt{\epsilon}.
\end{eqnarray}
As a result, we can find approximate expressions for the scalar spectral index $n_{s}$ and the tensor-to-scalar ratio $r$ as follows
\begin{eqnarray}
\epsilon&\simeq&\frac{1}{16}\left[\sqrt{4(1-n_{s})+\beta^{2}}-\beta\right]^{2},\\
r&\simeq&\left[\sqrt{4(1-n_{s})+\beta^{2}}-\beta\right]^{2}.\label{r-ns-bet-rel}
\end{eqnarray}
Note that, from Eqs. (\ref{rmX-rel}) and (\ref{r-ns-bet-rel}), we can express the parameter $\beta$ in terms of the scalar spectral index $n_s$ and the brane-vector mass $m_X$ as follows
\begin{eqnarray}
\beta\simeq 0.02\left[4(1-n_s)-958.48\left(\frac{\text{GeV}}{m_X}\right)^{7/4}\right]\left(\frac{m_X}{\text{GeV}}\right)^{7/8}.
\end{eqnarray}

The value of the inflaton at the end of the inflation corresponding to $\epsilon\simeq 1$ reads
\begin{eqnarray}
X_{\text{e}}\simeq\frac{1+n+2n\beta-\sqrt{1-n^{2}+4n(1+n)\beta+4n^{2}\beta^{2}}}{n(1+n)},
\end{eqnarray}
where 
$X_{\text{e}}\equiv\exp\left(-\phi_{\text{e}}/\bar{M}\right)$. Then, the number of e-folds is calculated by
\begin{eqnarray}
N_k=\frac{1}{2n\beta^2}\left[B(1-X_k,2+2n,-1)-B(1-X_{\text{e}},2+2n,-1)\right],\label{Nk-exp}
\label{eqn:nk}
\end{eqnarray}
where $B(z,p,q)=\int^z_0t^{p-1}(1-t)^{q-1}dt$ is the incomplete beta function.

\begin{figure}[t]
  \centering
  \includegraphics[scale=0.4]{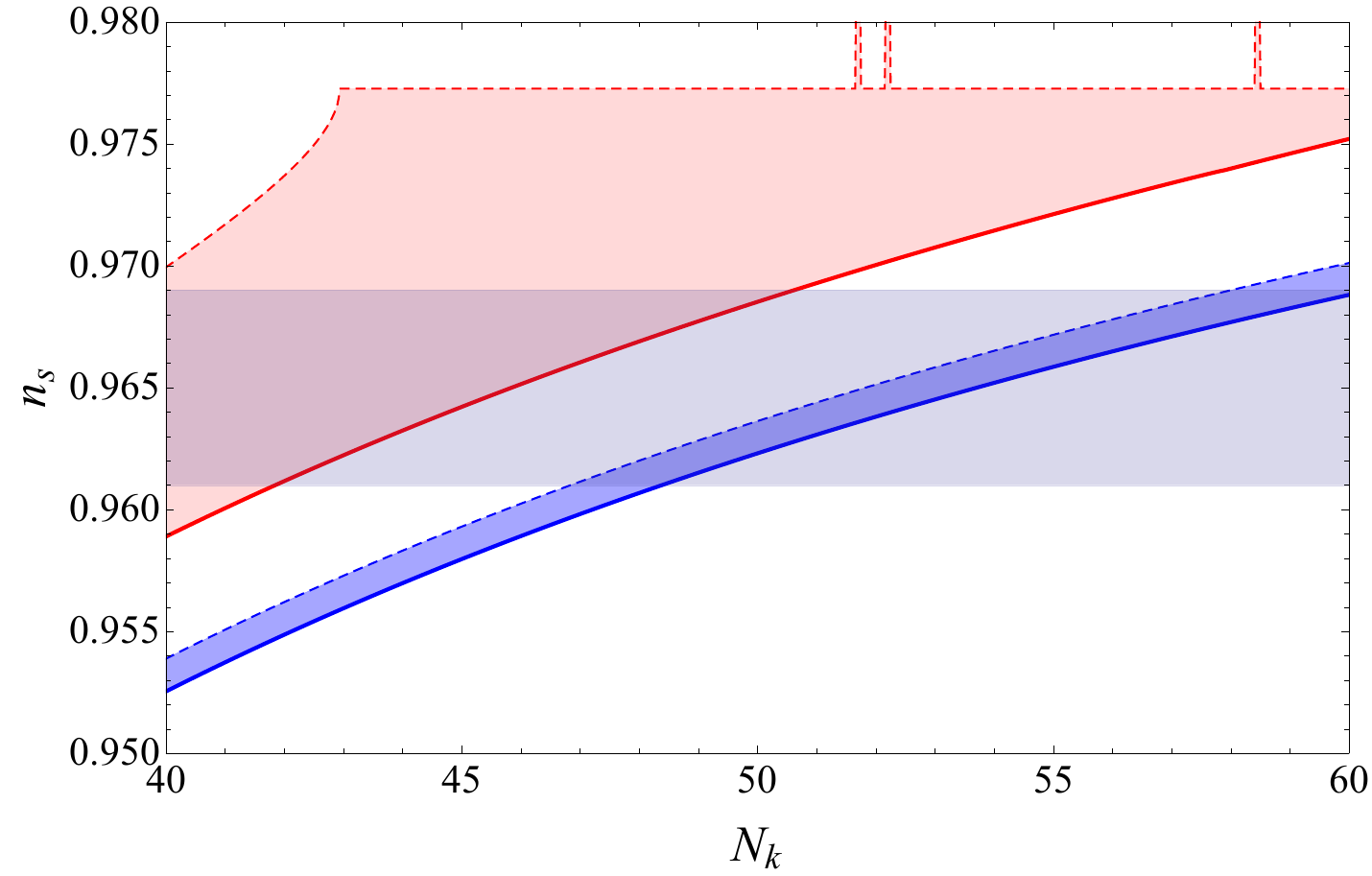}
  \caption{The number of e-folds $N_k$ versus the scalar spectral index $n_s$ for the $\alpha$-attractor potential inflation. The light red  and blue bands correspond to the brane-vector mass $m_X=350$ GeV and $1000$ GeV, respectively. The parameter $n$ takes the value from $1$ to $10$ in every band where the dashed and solid boundary curves correspond to $n=1$ and $n=10$, respectively. The gray band refers to the observation constraint on the scalar spectral index as $n_s=0.965\pm0.004$ (68$\%$ CL) \cite{Planck2020}.}
  \label{fig-Nk-ns-Emod}
\end{figure}

In order to find the relation between $N_k$ and $n_s$, we solve numerically Eq. (\ref{Nk-exp}) for various values of the number of e-folds. The result is shown in Fig. \ref{fig-Nk-ns-Emod} where $n$ is taken the values from 1 to 10 and the mass of the brane-vector DM is considered as $350$ GeV and $1000$ GeV.
We observe that the brane-vector DM would be consistent with the inflation which requires 50 - 60 e-folds for its high mass region. The low mass region of the brane-vector DM may be favorable to the inflation if the parameter $n$ in the inflation potential is large enough.

\begin{figure}[t]
  \centering
  \includegraphics[scale=0.4]{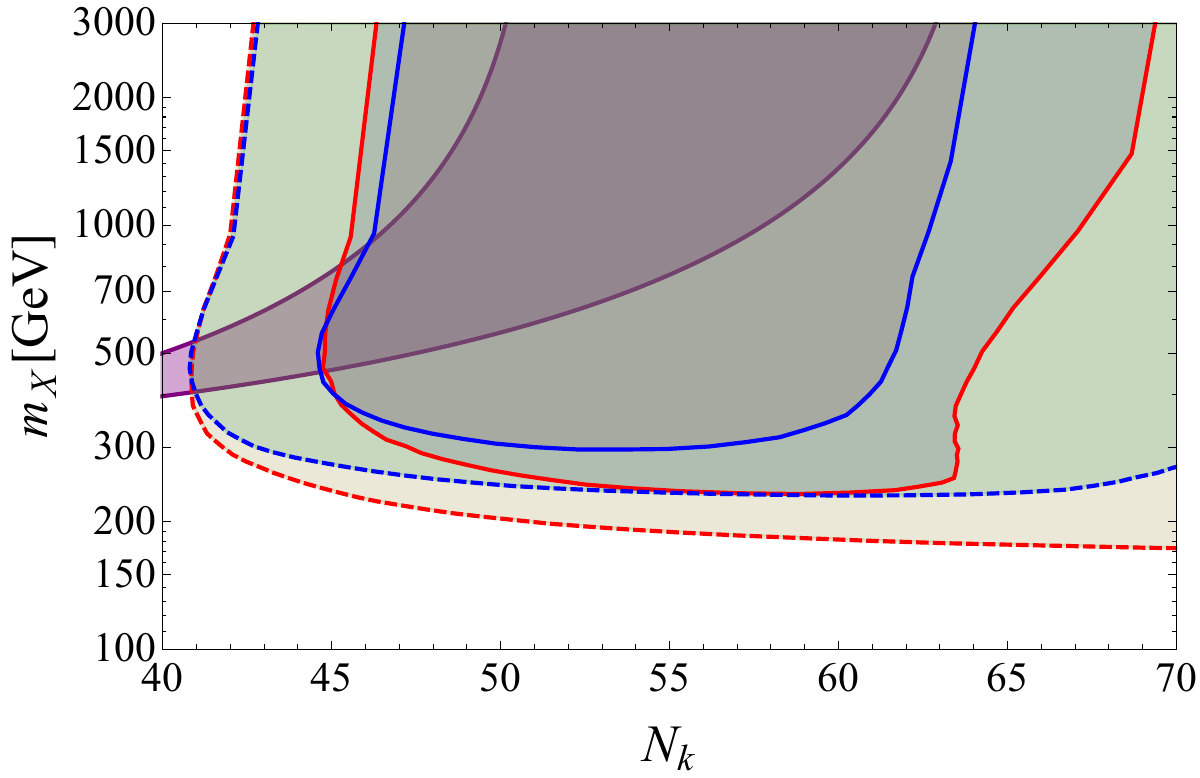}
  \includegraphics[scale=0.4]{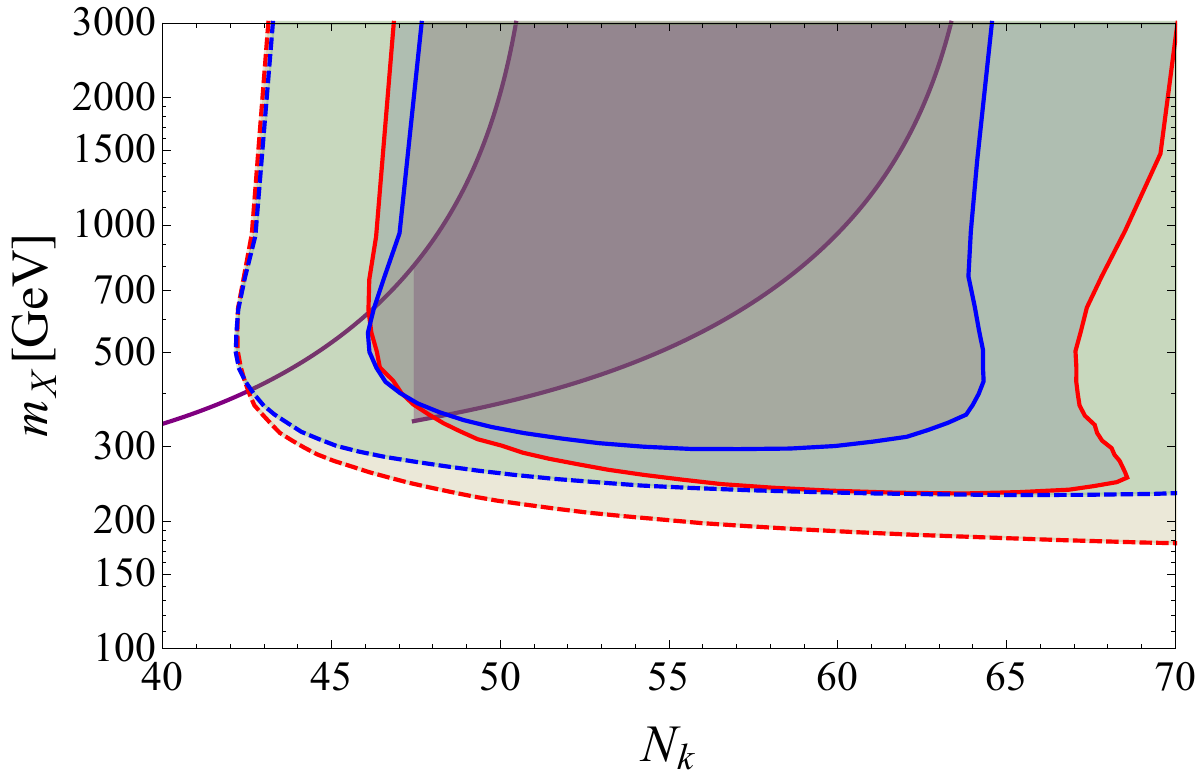}
  \caption{The number of e-folds $N_k$ versus the mass of the brane-vector DM with $n=1$ (the left panel) and $n=10$ (the right panel). Every purple band represents the prediction with the scalar spectral index taking the value from $0.965-0.004$ to $0.965+0.004$. The solid and dashed red contours correspond to the $68\%$ and $95\%$ CL Planck18 constraints for $n_s$ and $r$, respectively \cite{Planck2020b}. Whereas, the solid and dashed blue contours correspond to the $68\%$ and $95\%$ CL Planck18+BK15 constraints, respectively \cite{Planck2020b}.}
  \label{fig-Nk-mX-Emod}
\end{figure}

By solving numerically Eq. (\ref{Nk-exp}) and using Eq. (\ref{rmX-rel}), we obtain the relation between the number of e-folds $N_k$ and the mass of the brane-vector DM, as depicted in Fig. \ref{fig-Nk-mX-Emod}, with $n=1$ and $n=10$. Our prediction is shown by the purple bands with the scalar spectral index taking the value from $0.961$ to $0.969$. In adition, we show the $68\%$ and $95\%$ CL experimental constraints for $n_s$ and $r$ \cite{Planck2020b} corresponding to the (dashed) blue and (dashed) red contours.  
We see that the number of e-folds required by the inflation is satisfied in the high mass region of the brane-vector DM. Whereas, in order for the low mass region to meet this criterion, the parameter $n$ in the inflation potential needs to be large enough. These are completely consistent with what is observed in Fig. \ref{fig-Nk-ns-Emod}.

We arrive to investigate the reheating constraint for the brane-vector DM produced during the inflation. The connection between the parameters of the inflation and those of the reheating can be obtained by the following relation 
\begin{eqnarray}
\frac{\rho_{\text{re}}}{\rho_{\text{e}}}=e^{-3N_{\text{re}}(1+\omega_{\text{re}})},\label{re-endinf--enden}
\end{eqnarray}
where $\rho_{\text{re}}$ and $\rho_{\text{e}}$ are the energy density at the end of reheating (re) and the end of inflation, respectively, and $N_{\text{re}}$ is the number of e-folds during the reheating corresponding to the equation-of-state parameter denoted by $\omega_{\text{re}}$. The energy density at the end of reheating is determined in terms of the reheating temperature $T_{\text{re}}$ and the effective number of relativistic degrees of freedom $g_{\text{re}}$ at the reheating epoch as $\rho_{\text{re}}=\pi^2g_{\text{re}}T^4_{\text{re}}/30$. With the energy density at the end of inflation given by the inflation potential at the end of the inflation $V_{\text{e}}$ as $\rho_{\text{e}}\simeq3V_{\text{e}}/2$, we express the reheating temperature as follows
\begin{eqnarray}
T_{\text{re}}&=&\left[\left(\frac{45}{\pi^2g_{\text{re}}}\right)^3\frac{\pi^2A_s}{2}\right]^{1/12}\left(M^2_{\text{P}}m_X\right)^{1/3}\left(\frac{50.55\ \text{GeV}}{m_X}\right)^{7/48}e^{-\frac{3}{4}N_{\text{re}}(1+\omega_{\text{re}})}\nonumber\\
&&\times\left[\frac{n(n-2\beta)-1+\sqrt{1-n^{2}+4n(1+n)\beta+4n^{2}\beta^{2}}}{n(n+1)}\right]^{n/2}.\label{reh-temp}
\end{eqnarray}

In order to determine $N_{\text{re}}$, we use the relation $k/(a_0H_0)=a_kH_k/(a_0H_0)$ with $k=0.002$ Mpc$^{-1}$ being the pivot scale and $a_k$ ($H_k$) is the scale factor (the Hubble parameter) at the moment which the mode $k$ crosses the horizon, which leads to \cite{Liddle2003}
\begin{eqnarray}
\frac{k}{a_0H_0}=e^{-\left(N_k+N_{\text{re}}\right)}\frac{a_{\text{re}}}{a_{\text{mre}}}\frac{a_{\text{mre}}H_{\text{mre}}}{a_0H_0}\frac{H_k}{H_{\text{mre}}}.
\end{eqnarray}
From the conservation of entropy, we can find the ratio of $a_{\text{re}}$ to $a_0$ as follows
\begin{eqnarray}
 \frac{a_{\text{re}}}{a_0}=\left(\frac{43}{11g_{s,\text{re}}}\right)^{1/3}\frac{T_0}{T_{\text{re}}},   
\end{eqnarray}
where $g_{s,\text{re}}$ is the effective number of relativistic degrees of freedom that contribute to the entropy and $T_0$ is the present temperature of the universe. Then, the number of e-folds at the end of reheating is given by
\begin{eqnarray}
N_{\text{re}}&=&\frac{4}{1-3\omega_{\text{re}}}\left[67.38-N_{k}-\ln\left( \frac{k}{a_{0}H_{0}}\right)-\frac{1}{12}\ln g_{\text{re}} +\frac{1}{3}\ln \frac{g_{\text{re}}}{g_{s,\text{re}}}+\frac{1}{4}\ln\left(\frac{\pi^2A_s}{3}\right)\right.\nonumber\\
&&\left.+\frac{7}{16}\ln\left(\frac{50.55\ \text{GeV}}{m_X}\right)+\frac{n}{2}\ln\left(\frac{n(n+1)}{n(n-2\beta)-1+\sqrt{1-n^{2}+4n(1+n)\beta+4n^{2}\beta^{2}}}\right)\right].\label{Nefo-re}
\end{eqnarray}

\begin{figure}[ht]
  \centering
  \includegraphics[scale=0.31]{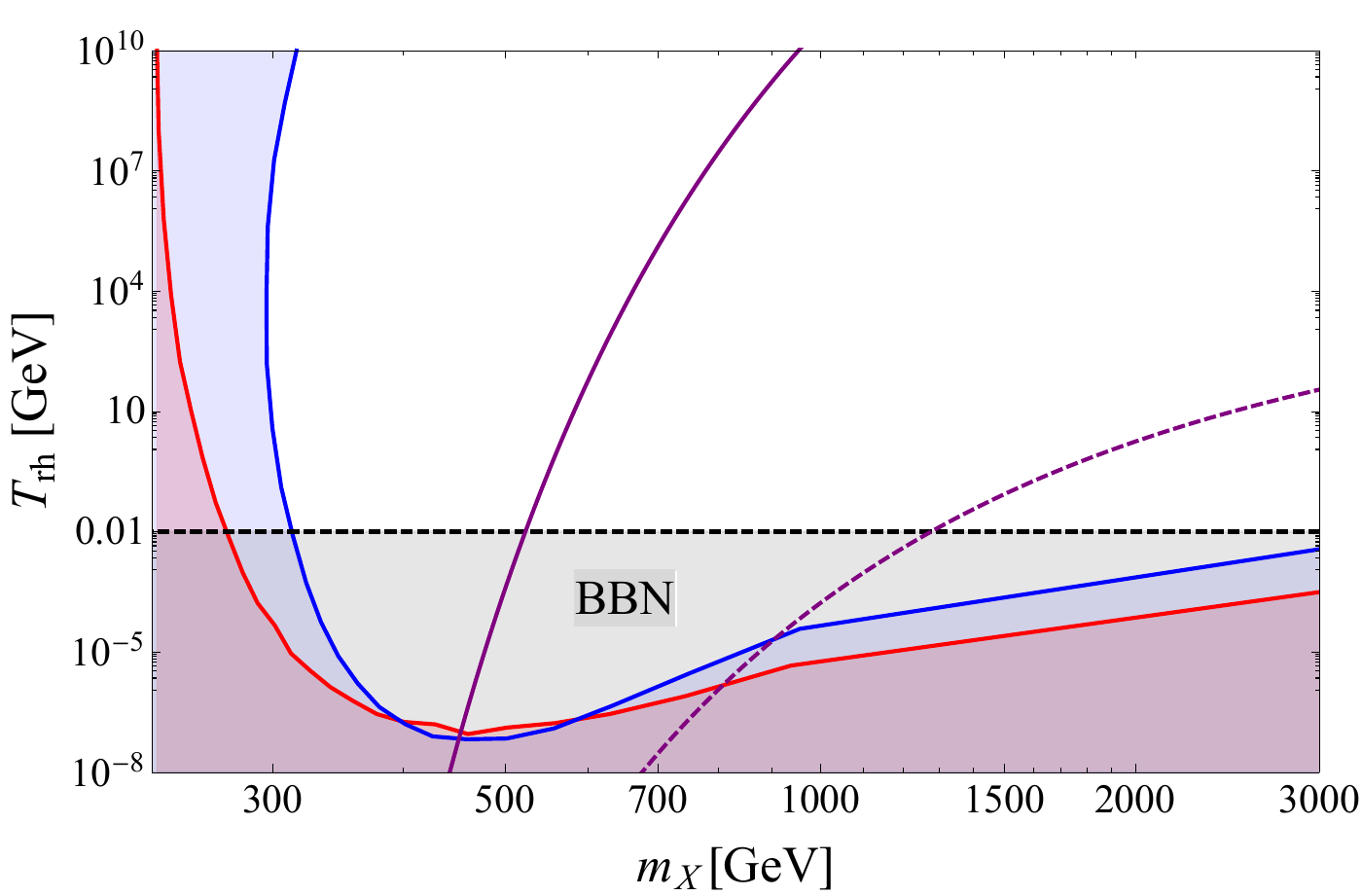}
   \includegraphics[scale=0.31]{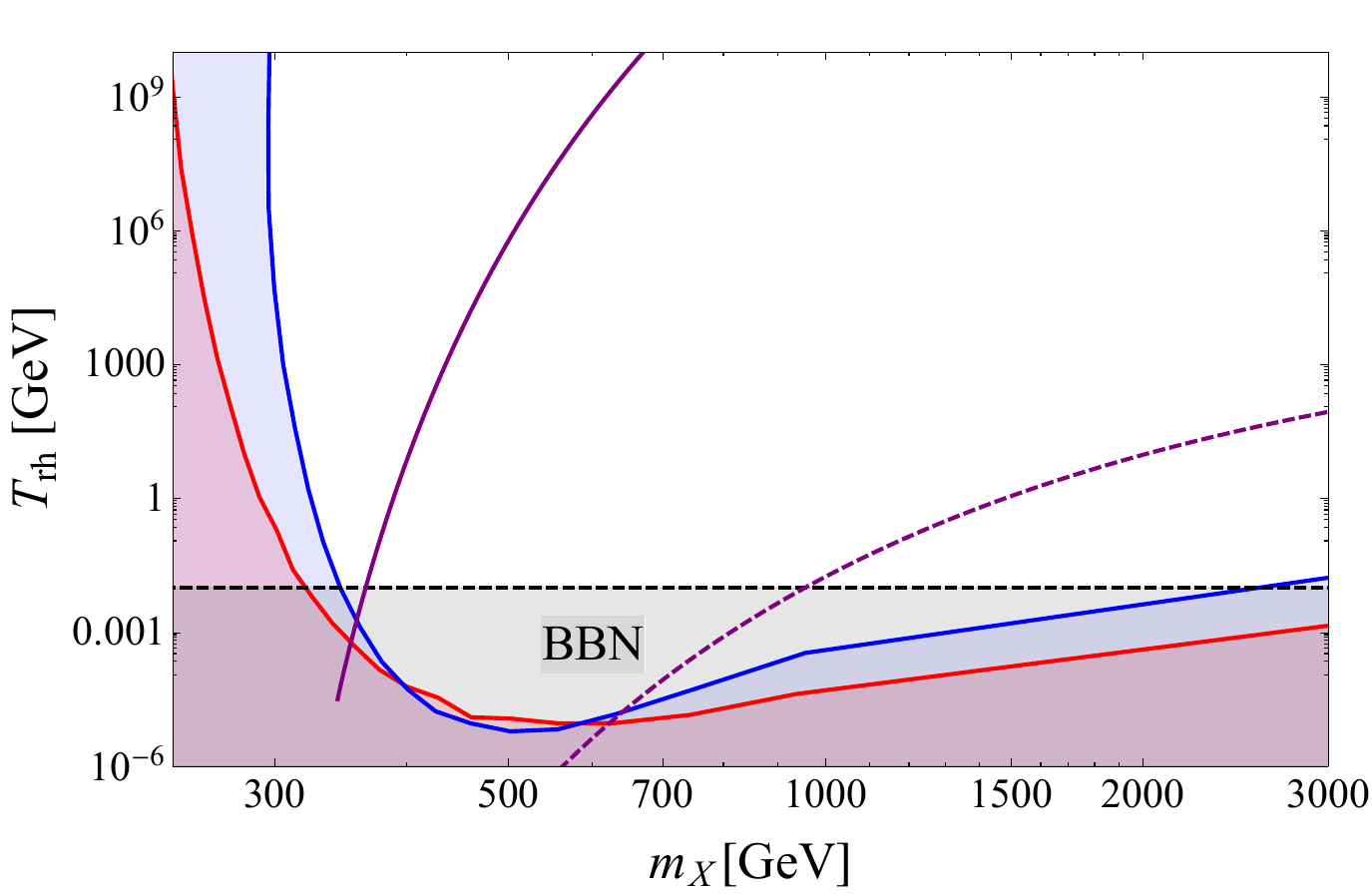}
  \caption{The reheating temperature versus the mass of the brane-vector DM for various values of the parameter $n$ with $\omega_{\text{re}}=0$. The left and right panels correspond to $n=1$ and $n=10$, respectively. The solid and dashed purple curves correspond to $n_s=0.969$ and $n_s=0.961$, respectively. The red and blue contours correspond to the $68\%$ CL Planck18 and Planck18+BK15 constraints, respectively. The grey region is excluded by the constraint of Big Bang Nucleosynthesis (BBN).}
  \label{fig-mX-Tre-Emod}
\end{figure}

From Eqs. (\ref{reh-temp}) and (\ref{Nefo-re}), we can find the relation between the reheating temperature $T_{\text{re}}$ and the mass of the brane-vector DM. In Fig. \ref{fig-mX-Tre-Emod}, we depict this relation for $n=1$ (the left panel) and $n=10$ (the right panel). The region is bounded by the solid and dashed purple curve represents our prediction with the $68\%$ CL value of the scalar spectral index as $n_s=0.965\pm0.004$.
We see that the present observation of $n_s$ leads to the large uncertainty for the prediction of the reheating temperature in terms of the mass of the brane-vector DM. The CMB experiments in the future which will be able to decrease the uncertainty of the $n_s$ measurement will provide a more accurate prediction for this relation.

\begin{figure}[ht]
  \centering
  \includegraphics[scale=0.32]{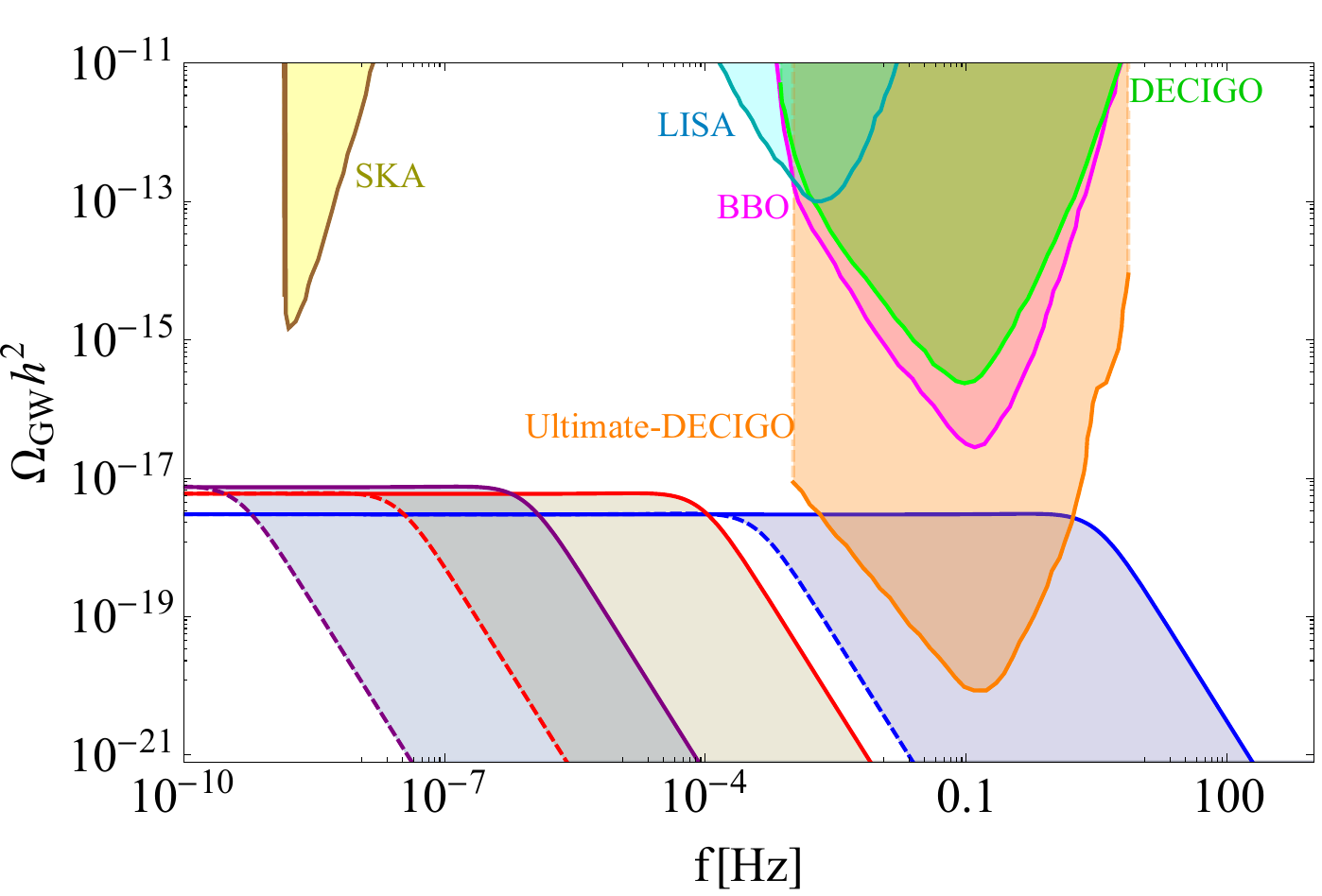}
   \includegraphics[scale=0.32]{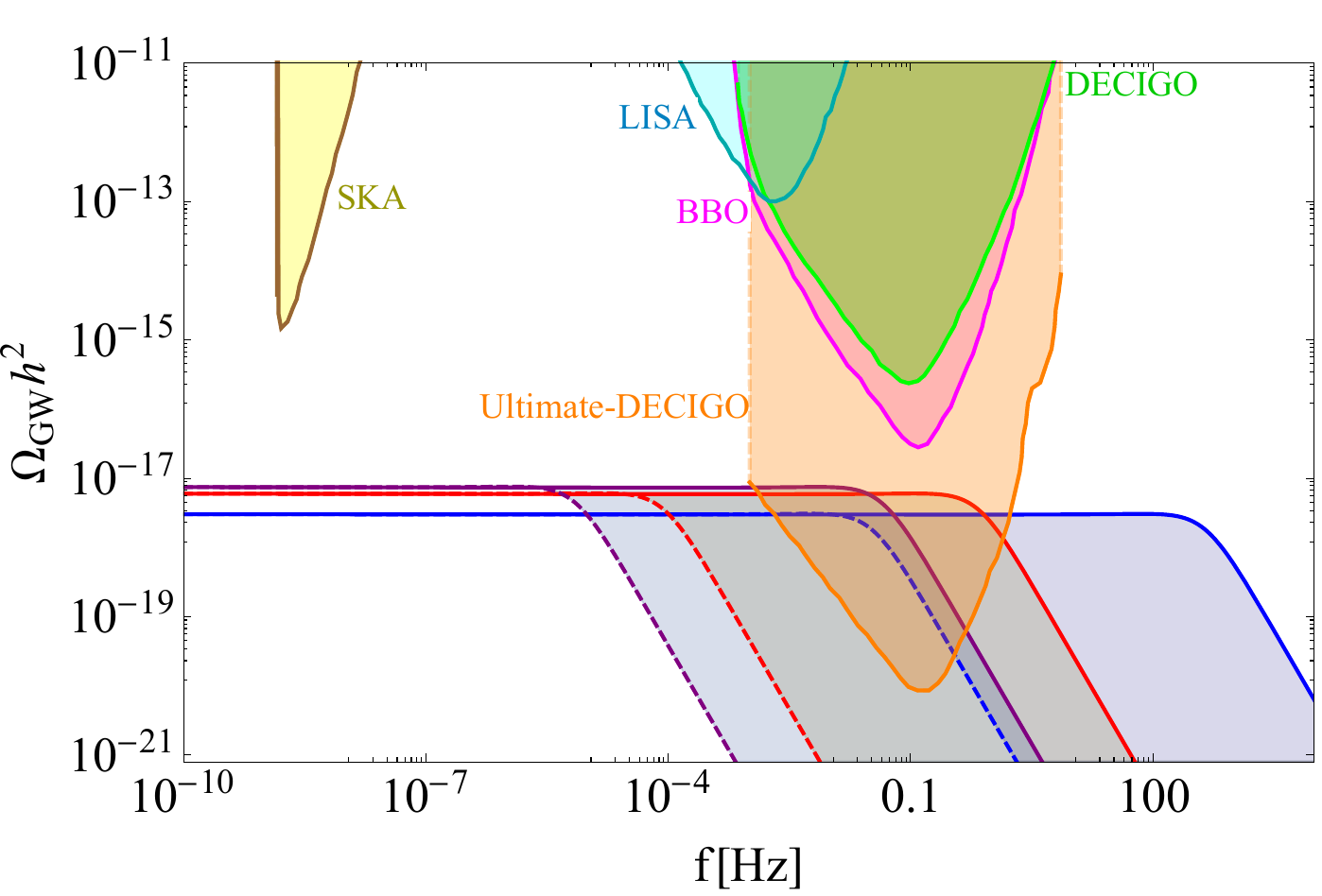}
  \caption{The spectrum of inflationary GWs as a function of the frequency for various values of the brane-vector mass and the parameter $n$. The scalar spectral index $n_s$ runs from $0.965$ (represented by the dashed blue, red, and purple curves) to $0.967$ (represented by the solid blue, red, and purple curves). The left and right panels correspond to $n=1$ and $n=10$, respectively. The bands bounded by the blue, red, and purple curves correspond to $m_X=1186$ GeV, $798$ GeV, and $702$ GeV, respectively.}
  \label{PGWs-fig}
\end{figure}

The spectrum of inflationary GWs \cite{Yokoyama2010}, which is a function of the frequency $f$, is connected to the brane-vector DM as
\begin{eqnarray}
\Omega_{\text{GW}}h^2\left(f\right)&\simeq&\frac{A_s}{3}\left(\frac{\pi\Omega_{\text{m}}hf}{a_0H_0}\right)^2\left(\frac{50.55\ \text{GeV}}{m_X}\right)^{7/4}\exp\left\{-\frac{1}{8}\left(\frac{50.55\ \text{GeV}}{m_X}\right)^{7/4}\log\left(\frac{f}{f_0}\right)\right\}\nonumber\\
&&\times\frac{g_{\text{in}}}{3.36}\left[\frac{3.9}{g_{s,{\text{in}}}}\right]^{4/3}\left[\frac{3j_1(z_k)}{z_k}\right]^2T^2_1\left(f/f_{\text{eq}}\right)T^2_2(f/f_{\text{re}}),
\end{eqnarray}
where ``in" denotes the moment that the mode $k$ crosses the horizon, $j_1(z)$ is the spherical Bessel function of the first kind, $z_k\equiv4\pi f/(a_0H_0)$, the functions $T_{1,2}(x)$ describe the change of the primordial GW spectrum at the frequency $f_{\text{re}}\simeq0.26\text{Hz}\times T_{\text{re}}/(10^7\text{GeV})$ \cite{Kuroyanagi2011} due to the presence of the reheating and are given as follows
\begin{eqnarray}
T_1(x)&=&\left(1+1.57x+3.42x^2\right)^{1/2},\nonumber\\
T_2(x)&=&\left(1-0.32x+0.99x^2\right)^{-1/2}.
\end{eqnarray}
In Fig. \ref{PGWs-fig}, we show the prediction of the primordial GW spectrum for various values of the brane-vector mass and the parameter $n$. In addition, we include the projected sensitivities of the different GW observations, represented by the colored regions, which are the future SKA \cite{SKA}, the space-based LISA \cite{LISA}, BBO \cite{BBO}, DECIGO \cite{DEC}, and Ultimate-DECIGO \cite{UDEC}.
We observe that Ultimate-DECIGO can probe the reheating temperature and the brane-vector DM in its allowed mass region if the parameter $n$ in the inflation potential is large enough. On the contrary, if $n$ is low, Ultimate-DECIGO only allows us to probe the low mass region of the brane-vector DM. In this sense, the observation of the primordial GWs produced during the inflation provides a potential way to probe the extra-dimensions and branes: the reheating temperature and the mass of the brane-vector DM which are determined by the spectrum of the primordial GWs are compared to the prediction which is given in Eqs. (\ref{reh-temp}) and (\ref{Nefo-re}).

\subsection{Natural potential}
We study the connection of the brane-vector DM to the inflation with the natural potential or the axion potential \cite{Freese1990} which appears naturally in the compactifications of M/string theory \cite{Svrcek2006} and provides a plausible solution to the strong CP problem \cite{Peccei1977}. The inflation potential in the natural inflation is given as follows
\begin{eqnarray}
V(\phi)=\Lambda^4\left[1-\cos\left(\frac{\phi}{f}\right)\right],\label{nat-pot}
\end{eqnarray}
where $f$ is a parameter. Using Eq. (\ref{sl-rol-pars}) with the inflation potential (\ref{nat-pot}), we find the slow-roll parameters as follows
\begin{eqnarray}
\epsilon = \gamma^{2}\frac{1+\cos(\phi/f)}{[1-\cos(\phi/f)]^2}\\
\eta = \gamma^{2}\frac{\cos(\phi/f)}{[1-\cos(\phi/f)]^{2}}
\end{eqnarray}
where
\begin{eqnarray}
\gamma \equiv \frac{M_{\text{P}}^{2}m_{X}}{f\Lambda^{2}}.
\end{eqnarray}
At the horizon crossing of the mode $k$, corresponding to $ \epsilon\ll 1 $, we find the value of the inflaton field in terms of the spectral index $n_s$ and the brane-vector mass as follows
\begin{eqnarray}
\cos\left(\frac{\phi_k}{f} \right)\simeq -1 + \frac{4}{25\gamma^{2}}\left[5(1-n_s)-\gamma\left(\gamma+\sqrt{(5-5n_s)/2+\gamma^2}\right)\right]\equiv X_k,
\end{eqnarray}
where the parameter $\gamma$ can be expressed as
\begin{eqnarray}
\frac{\gamma}{\sqrt{2}}\simeq\frac{1-n_{s}-300\left(\text{GeV}/m_X\right)^{7/4}}{\sqrt{1-n_{s}-240\left(\text{GeV}/m_X\right)^{7/4}}}.
\end{eqnarray}

The number of e-folds is
\begin{eqnarray}
N_k=\frac{1}{\gamma^{2}}\left[X_{k}-X_{f} - 2\log\left( \frac{1+X_{k}}{1+X_{f}}\right)  \right],  
\end{eqnarray}
where $X_f\equiv\cos(\phi_f/f)$ with $\phi_f$ being the value of the inflaton field at the end of inflation which is determined by the condition $ \epsilon\simeq 1 $ leading to
\begin{eqnarray}
X_{f}\simeq 1+\frac{\gamma}{2}\left(\gamma-\sqrt{\gamma^{2}+8}\right).   
\end{eqnarray}

\begin{figure}[ht]
  \centering
  \includegraphics[scale=0.4]{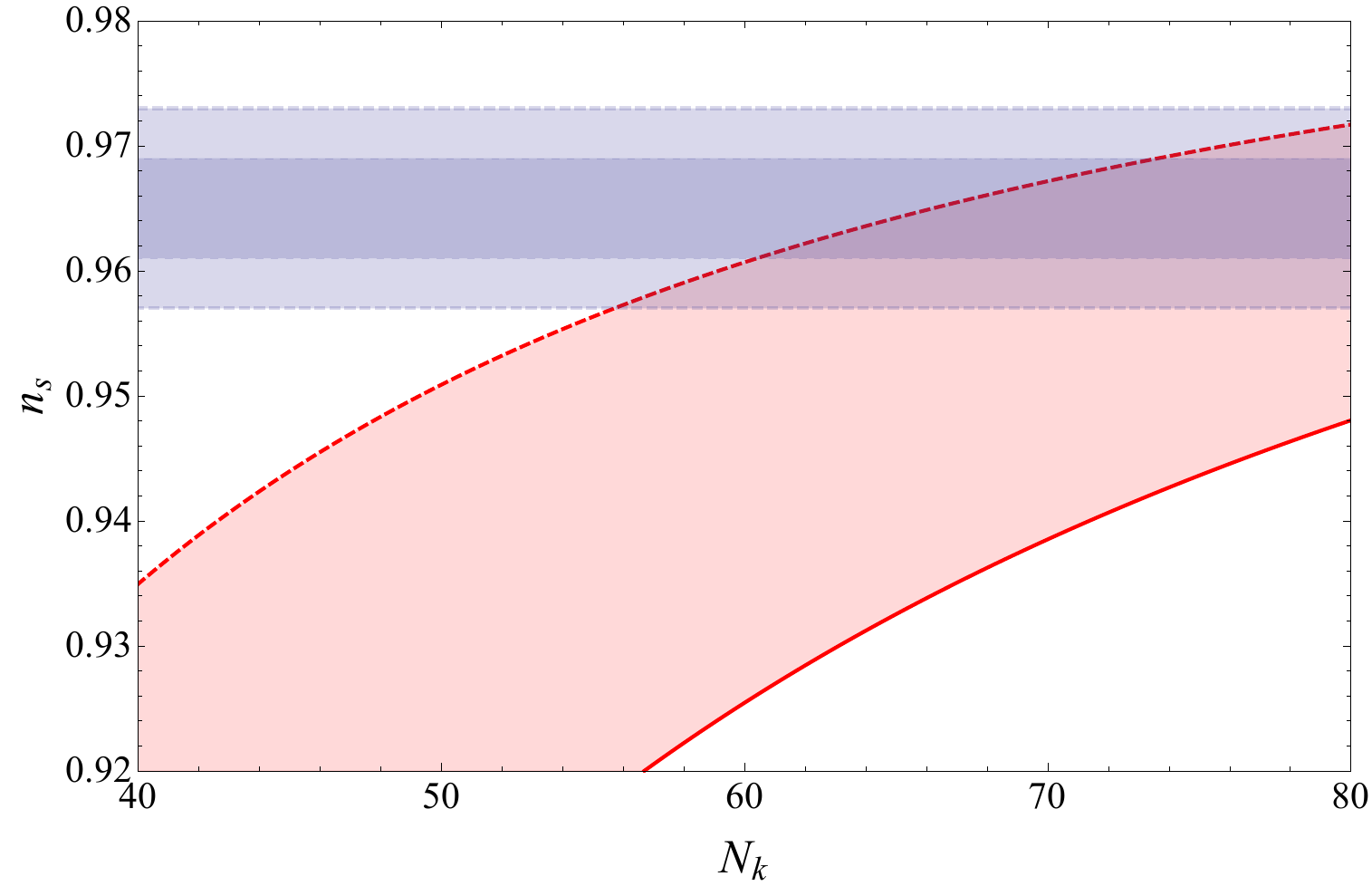}
  \caption{The number of e-folds $N_k$ versus the scalar spectral index $n_s$ in the natural inflation for various values of the brane-vector mass from $m_X=280$ GeV (dashed red curve) to $m_X=1000$ GeV (solid red curve). The dark and light gray bands correspond to the value of $n_s$ at 68$\%$ CL and $95\%$ CL, respectively.}
  \label{fig-Nk-ns-natpot}
\end{figure}

In Fig. \ref{fig-Nk-ns-natpot}, we show the dependence of the scalar spectral index $n_s$ on 
the number of e-folds $N_k$.
We can see that the brane-vector DM is only consistent with the natural inflation in the $2\sigma$ region for the value of $n_s$ and its low mass region around 280 GeV. However, as we see above, the low mass region of the brane-vector DM is hard to detect in the spectrum of the primordial GWs in the near future.

\section{\label{conclu}Conclusion}

In the present paper, we have explored dark matter and its connection to the inflationary observations and the spectrum of the primordial gravitational waves in a scenario in which our observable universe is realized as a flexible 3-brane that could oscillate in a five-dimensional bulk spacetime compactified on a circle. (Such a scenario can be considered as a low-energy limit of ten-dimensional string theory which is reduced on a five-dimensional internal manifold.) The off-diagonal components of the bulk metric that is the gauge field corresponding to the U(1) isometry of the internal space would absorb the scalar mode describing the fluctuation of the 3-brane in the bulk. As a result, this leads to a massive vector field living on the 3-brane, so-called the brane-vector. Because the coupling of the brane-vector is strongly suppressed by the observed Planck scale, the observed relic abundance of the brane-vector dark matter is essentially due to the production of its longitudinal mode by the quantum fluctuations in the inflationary stage. In this way, the tensor-to-scalar ratio is determined in terms of the brane-vector mass or the brane tension. In addition, the mass of the brane-vector enters into the Friedmann equation describing the evolution of the brane universe, which is very sensitive with respect to the region of the high energy density such as the inflation scale. Therefore, we have studied the imprints of the brane-vector dark matter in the inflationary observables and the spectrum of the primordial gravitational waves. This can open a promising observation window to detect the extra-dimensions and branes which play a fundamental role in string/M theory.

\end{document}